\documentclass[twocolumn,pra,aps]{revtex4}
\usepackage{graphicx}

\begin{document}

\title{Quantum randomness can be controlled by free will
\\--a consequence of the before-before experiment}

\author{Antoine Suarez}
\address{Center for Quantum Philosophy, P.O. Box 304, CH-8044 Zurich, Switzerland\\
suarez@leman.ch, www.quantumphil.org}

\date{January 19, 2009}

\begin{abstract}

The before-before experiment demonstrates that quantum randomness can be controlled by influences from outside spacetime, and therefore by immaterial free will. Rather than looking at quantum physics as the model for explaining free will, one should look at free will as a primitive principle for explaining why the laws of Nature are quantum.\\

\footnotesize\emph{Key words}: Quantum randomness, free will, before-before experiment, agency from outside spacetime.

\end{abstract}


\maketitle

\section {The ambiguity of the term 'randomness'}

The assumption that human behavior is not \emph{completely} determined by the past plays a key role in the way we behave in daily life and organize society through law. When I typewrite this article, I assume that I am governing the movements of my fingers through my free will. Accordingly I claim to be the author of the article and to express original thoughts, which are not \emph{completely} pre-determined at the beginning of the Universe. Anyone who claims for the right ``to choose how to live his life'' should coherently exclude any explanation of his brain using \emph{only} deterministic causality, be it in terms of genes, chemicals or environmental influences.

In this sense, the principle of freedom conflicts with the deterministic description of classical physics. The philosopher Immanuel Kant vividly experienced this conflict in his own intellectual life, and declared: ``it cannot be alleged that, instead of the laws of nature, laws of freedom may be introduced into the causality of the course of nature. For, if freedom were determined according to laws, it would be no longer freedom, but merely nature.''\cite{kant} This conclusion was inescapable within the deterministic science of Kant's time.

By contrast today's quantum physics assumes events which are not completely determined by the past, and cannot be explained by means of observable causes alone. In this sense this theory offers a description of the world which does not exclude free-willed agency in principle.

\begin{figure}[t]
\includegraphics[width=80 mm]{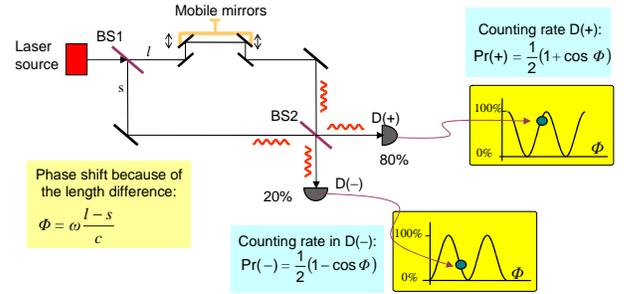}
\caption{Single particle interference is the entry to the Quantum World. Laser light of frequency $\omega$ emitted by the source is either transmitted or reflected at each of the beam-splitters (half-silvered mirrors) BS1 and BS2; therefore the light can reach the detectors D(+) and D($-$) by the paths $l$ and $s$; the path-length $l$ can be changed by the experimenter. With sufficiently weak intensity of light, only one of the two detectors clicks: either D($+$) or D($-$) (photoelectric effect). Nevertheless, Nature calculates the counting rates of each detector Pr($+$) and Pr($-$) taking account of the length of the two paths $l$ and $s$ (interference effect). According to quantum mechanics, which detector clicks in a single run is decided by a true choice (on the part of Nature) when the information about the two paths reaches the detectors: There is nothing in the universe allowing us to predict which alternative will happen.}
\label{f1}
\end{figure}

Suppose the quantum experiment sketched in Figure \ref{f1}. When one works with sufficiently weak intensity of light, then only one of the two detectors clicks: either D($+$) or D($-$) (photoelectric effect). Nevertheless, for calculating the counting rates of each detector one must take into account information about the two paths leading from the laser source to the detector: The probability that one detector clicks in a large series of runs can be exactly predicted for each detector, and depends deterministically on the path-length difference (interference effect). According to quantum mechanics, which detector clicks in a single run in the experiment of Figure \ref{f1} is decided by a true choice (on the part of Nature) when the information about the two paths reaches the detectors. Which detector clicks is not only unpredictable for us because we don't yet know the formula connecting the past with the present and the present with the future, but it is unpredictable in principle because such a formula doesn't exist at all.

The event that D($+$) clicks and D($-$) does not click is often said to be ``a genuinely random event'' \cite{vv}. 'Random' means here that Nature's decision about which detector clicks, though it has some roots in the past, is \emph{not completely determined by the past}.

A frequent objection against the possible relevance of quantum physics for the question of free will is that quantum non-deterministic randomness \emph{excludes} the possibility of order and control, and therefore free will. Thus one states for instance that: ``In the end, however, it is clear that neither determinism nor randomness is good for free will. If nature is fundamentally random, then outcomes of our actions are also completely beyond our control: randomness is just as bad as determinism''\cite{vv}. Here the term 'randomness' refers to \emph{'exclusion of order and control'}, and not \emph{'not completely determined by the past'}.

This paper aims to show that quantum physics does not entail the presumed incompatibility of quantum randomness with order and control. Section II argues on the basis of the before-before experiment that quantum randomness can be controlled by unobservable influences from outside spacetime and, therefore, is compatible with freedom in principle: Both quantum randomness and free will, refer to agency which is not exclusively determined by the past. Section III answers a number of other objections. Section IV concludes by stating that one should look at free will as a primitive principle for explaining why the laws of Nature are quantum, instead of looking at the quantum as the model for explaining free will.

\section {Quantum randomness can be controlled by unobservable influences from outside spacetime}

\begin{figure}[t]
\includegraphics[width=80 mm]{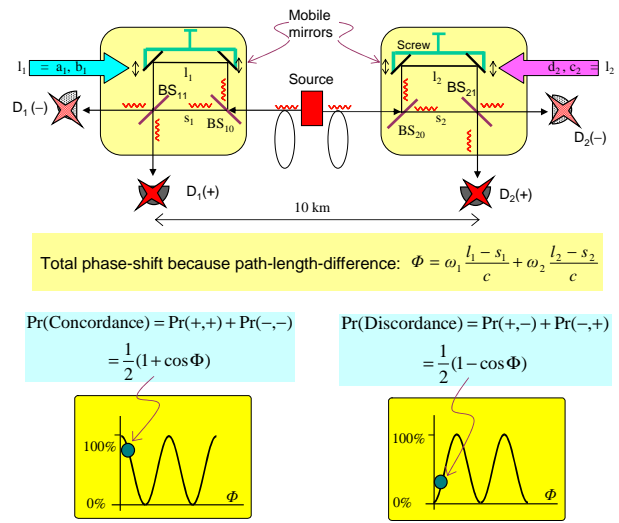}
\caption{In entanglement experiments, local random events can be influenced from outside spacetime to produce nonlocal patterns (correlations). The source emits photon pairs. Photon 1 (frequency $\omega_{1}$) enters the left interferometer through the beam-splitter BS$_{10}$ and gets detected after leaving the beam-splitter BS$_{11}$, and photon 2 (frequency $\omega_{2}$) enters the right interferometer through the beam-splitter BS$_{20}$ and gets detected after leaving the beam-splitter BS$_{21}$. The detectors are denoted D$_{1}(a)$ and D$_{2}(b)$ ($a, b\in\{+,-\}$). Each interferometer consists in a long arm of length $l_{i}$, and a short one of length $s_{i}$, $i\in\{1,2\}$. Frequency bandwidths and path alignments are chosen so that only the coincidence detections corresponding to the path pairs: $(s_{1},s_{2})$ and $(l_{1},l_{2})$ contribute constructively to the correlations, where $(s_{1},s_{2})$ denotes the pair of the two short arms, and $(l_{1},l_{2})$ the pair of the two long arms. Depending on the value of the phase $\Phi$, the quantity $Pr(a, b)$ oscillates between the situation of perfect correlation $Pr(a=b)=1$, and that of perfect anticorrelation $Pr(a=b)=0$. However, $Pr(a)$ for photon 1 alone is always 50\%, and $Pr(b)$ for photon 2 alone is 50\% too.
Bell experiments, using two different values of $l_{1}$ and two different values of $l_{2}$, demonstrate that the correlations violate the locality criteria (Bell's inequalities).
The before-before experiment demonstrates that nonlocal correlations have their roots outside of space-time. In this experiment the beam-splitters BS$_{11}$ and BS$_{21}$ are in motion in such a way that each of them, in its own reference frame, is first to select the output of the photons (before-before timing). Then, each outcome should become independent of the other, and the nonlocal correlations should disappear. The result was that the correlations doesn't disappear, and therefore are independent of any time-order.}
\label{f2}
\end{figure}

Quantum mechanics predicts correlated outcomes in space-like separated regions for experiments using 2-particle entangled states like the represented in Figure \ref{f2}. Suppose one of the measurements produces the value $a$ ($a\in\{+,-\}$), and the other the value $b$ ($b\in\{+,-\}$). According to quantum mechanics the probability $Pr(a, b)$ of getting the joint outcome $(a, b)$ depends on the choice of the phase parameters characterizing the paths or channels uniting the source and the detectors; depending on the value of the phases, the quantity $Pr(a, b)$ oscillates between the situation of perfect correlation $Pr(a=b)=1$, and that of perfect anticorrelation $Pr(a=b)=0$. However, $Pr(a)$ for photon 1 alone is 50\%, and $Pr(b)$ for photon 2 alone is 50\% too.

There are two alternative ways of explaining these quantum correlations, depending on whether one assumes the freedom of the experimenter as an axiom or not. In the context of the experiment illustrated in Figure 2, this axiom states that in choosing the values of the path-lengths \emph{l} the physicists are not \emph{completely} determined by the past.

I would like to insist that the ``freedom of the experimenter'' is not something one can settle by experiment or computing, but is a matter of principle: you can either choose or reject it.

If you reject it, then you can explain things in a entirely (local or nonlocal) deterministic way:

You can for instance assume like Gerard 't Hooft sort of local ``super-deterministic'' conspiracy in Nature: Things are so contrived that the physicist unconsciously set path-lengths (Figure 2) which fit to the properties the particles carry in order to produce the correlations \cite{hooft07}.

Or you can choose the ``Many Worlds'' picture, which also is fully local deterministic \cite{jb87, vv}: Everything that can happen does in fact happen, but in different worlds, and thus, actually, there is no violation of Bell inequalities. This violation originates from the feeling that you are 'someone' living in a particular world. But this feeling is an illusion, because you cannot know ``which 'you' is you, and which 'you' is a copy''. It is a merit of this picture to show that determinism implies the loss of personal identity.

Or finally you can assume a picture in the sense of David Bohm \cite{db}, invoking influences which are both faster than light (nonlocal) and deterministic. However in this case determinism is rather fictitious because the time one assumes does not correspond to any real clock \cite{as0708}. Moreover, by choosing ``Super-determinism'' or ``Many Worlds'' you get rid of the spooky nonlocality for the same price.

If you accept the axiom of freedom on the part of the experimenter, then you are necessarily led to accept nonlocal influences between the apparatuses on both parts of the setup of Figure \ref{f2}, i.e., influences which cannot be explained by signals traveling with $v\leq c$, as Bell experiments show \cite{jb87, exp}. Moreover, the before-before experiment \cite{szsg} rules out the belief that physical causality necessarily is time ordered, so that an observable event (the effect) always originates from another observable event (the cause) occurring before in time. This experiment demonstrates that the nonlocal correlations have their roots outside of spacetime \cite{as0705}.

Additionally, in quantum entanglement experiments (Figure \ref{f2}) local randomness and nonlocal order appear inseparably united: A random event event A is nonlocally correlated according to statistical rules to another random event B. This means that non pre-determined (``genuinely random'') local events can be controlled to generate nonlocal patterns (correlations) and, as the before-before experiment shows, the control happens from outside spacetime. The quantum correlations do not result automatically from the real properties the particles carry and the real settings they meet: they cannot be explained by any history in the spacetime \cite{as0705, cb}. In other words, the quantum correlations require decisions that take place when measurement happens, and establish ordered outcomes.

The before-before experiment supports David Hume's view that causation in time is an illusion of our intuition. The true causes are invisible and act from outside spacetime. We know these causes through the visible effects which they produce --the nonlocal correlations.

\begin{figure}[t]
\includegraphics[width=80 mm]{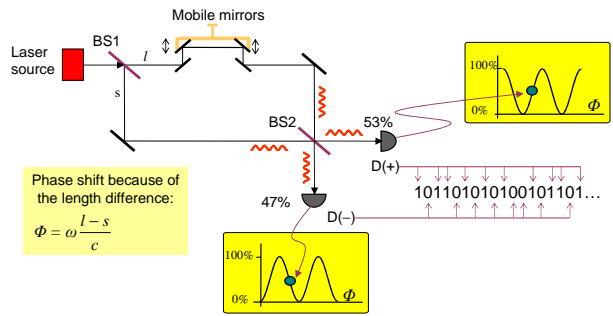}
\caption{In quantum interference, a single outcome, either '$+$' or '$-$', represent a bit of information, and a large series of outcomes builds a string of bits 1 and 0, a piece of information. The path-length difference determines the statistical distribution of the outcomes for large series, say for instance: 53\% '$1$' and 47\% '$0$'. Quantum physics imposes nothing regarding the order in which the single outcomes occur, provide the statistical distribution is fulfilled for a ``long'' series of bits. It is possible in principle that an unobservable mental variable (a free willed intellect) influences the order of the bits during a time, and encodes a message in the string.}
\label{f3}
\end{figure}

The result that immaterial agency controls quantum randomness to produce correlations (order), can open the road towards an explanation of how a mind may purposefully generate more sophisticated pieces of information. Consider the interference experiment of Figure \ref{f3}. A single outcome, either '$+$' or '$-$', represent a bit of information, and a series of outcomes (a bit string) build a piece of information. Quantum physics requires that long series of  outcomes fulfill a statistical distribution imposed mainly by the parameter of the path-length difference. Quantum physics imposes nothing regarding the order in which the single outcomes occur, and does not establish how long a series must be, to be considered ``long''.

Thus, it is possible in principle that an unobservable mental variable (a free-willed intellect) influences purposefully the order of the bits during a time, and encodes a message in the string. In any case, further purposeless outcomes would restore the outcomes' distribution the parameter impose.

This means that quantum indeterminism can be used purposefully, in accord with Anton Zeilinger's view of the ``two freedoms'': there is the freedom on the part of Nature, additionally to the freedom on the part of the experimenter \cite{az}.

As said in Section I, the postulate of \emph{free-will or choice on the part of Nature} was tacitly assumed in the picture of the ``collapse of the wave function'' in experiments like that of Figure \ref{f1}. Nevertheless only after the before-before experiment the choice on the part of Nature can be considered demonstrated.

John Conway and Simon Kochen state: ``If indeed there exist any experimenters with a modicum of free will, then elementary particles must have their own share of this valuable commodity.''\cite{coko} I think that on the basis of the before-before experiment one can conclude more precisely: If I claim to be free, then I have also to accept that there is free will controlling the outcomes in quantum experiments. Not that the ``particles'' are free, but the mind deciding (from outside the spacetime) which detector clicks.

If by 'random' one means events that are not completely determined by the past, one can very well consider that quantum randomness and free will have the same origin. Your free will is for me as unpredictable as the best random number generator. One can interpret the before-before experiment in terms of ``nonlocal randomness'' or ``nonlocal free-will'' as well. Nicolas Gisin states: ``the same randomness manifests itself at several locations'' \cite{ng}). I state: ``a mind influences local randomness from outside spacetime to produce nonlocal order.'' I think the two statements are equivalent.

\section {Objections}

In the preceding Section I have stressed that determinism excludes free will, whereas quantum physics allows us to describe a world where freedom is possible. I now answer several objections against this view.

\subsubsection {Were all those who believed in free will before the arrival of quantum physics making an irrational
assessment?}

As referred to above in Section I, the philosopher Kant was fully aware of the contradiction between freedom and deterministic science.

In the past, supporters of the deterministic view of science who believed in free will, were certainly making a self-contradictory assessment though they may have overlooked it. Today, the powerful and successful search for neural correlates of human cognition and behavior does not longer permit to hide the deterministic threat, and Kant's dilemma is becoming especially acute. Some years ago the German neuroscientist Wolf Singer reformulated the dilemma as follows: ``We experience ourselves as free mental beings, but the scientific view does not admit any room for a mental agent like the free will, which influences neurons and produces actions [...] In my eyes this conflict cannot be solved for the time being. Both descriptions can be shared together even by researchers of the brain: when I observe the brain I cannot find any evidence of a mental agent like the free will or the own responsibility --nevertheless when I get home in the evening I hold my children responsible for their actions if they have done any nonsense.''\cite{ws00}

Like Singer, most of those who claim that classical neural-network computations are entirely sufficient to deal
with the problems encountered by nervous systems \cite{kh}, definitely declare that they are ``for freedom''. One may try to solve the incompatibility between freedom and deterministic neuroscience by claiming that freedom is ``a moral or philosophical issue''. Nevertheless, as Singer suggests, this explanation is not satisfactory. Indeed, science tries to explain the brain, and the wish for freedom undoubtedly happens in the brain \cite{gr99}: if the brain processes are actually completely determined by the past, it is not clear how they can produce such a wish, i.e. the wish for ``not being completely determined by the past''.

Placed in front of the conflict between freedom and deterministic neuroscience two rational positions are possible: Either you claim that freedom is an illusion. Or you consider that deterministic neuroscience is not the last word about the brain and has to be superseded by a neuroscience admitting processes which are not completely determined by the past.

Singer himself seems now more open to this second possibility, and uses to begin his talks with the warning: ``Let us be careful. Perhaps there are in the world still things that, if we discover them, will turn the world inside out like quantum physics did with classical physics.''\cite{ws06}

I think these words point to the right direction: In any case, we cannot solve the problem with a description on the basis of classical physics. A number of neuroscientist considers that quantum mechanics is not appropriate either, and say that we need a ``new physics'' to understand ``exotic'' brain states like mind and consciousness, ``a physics that does not yet exists, but obviously does not violate the current laws of Nature.'' \cite{gr99}

Even agreeing that quantum physics must be completed in a number of points, I consider that the skepticism about the compatibility of quantum physics with free will originates from misunderstandings: As argued above, the free will of the experimenter is not a theorem deriving from the quantum, but an axiom one can freely accept or reject. If one assumes the axiom, then the before-before experiment demonstrates that quantum randomness can be controlled by immaterial free will. As soon as these consequences are understood I am confident that the skepticism will disappear: Quantum physics provides a balanced mixture of deterministic statistical rules and non-deterministic single events to account for a world where freedom is possible.

\subsubsection {In the lab we do not meet quantum devices printing out poems, scientific papers, or any meaningful message.}

Such a possibility is not forbidden at all by quantum physics, provided statistical rules for ``long'' series of outcomes are not violated. Accordingly, I think that in daily life we meet plenty of ``quantum devices'' (human brains) communicating efficiently with each other.

Many features of my brain's physiology are susceptible to deterministic description in terms of observable causal chains (the metabolism involved in the arousal potentials triggering bodily movements, for instance, follows the usual physical conservation laws). Additionally, the physiological measurable parameters of the brain are fixed by a number of factors (genetic, epigenetic and environmental ones), and in particular by sensorial impressions. These parameters determine the statistical distribution of the brain outcomes for large series at a given time. During certain periods of time a brain produces meaningful pieces of information (speech, text, musical composition, painting...). During other periods of time (while sleeping and even during many wake periods) the brain produces incoherent random signals.

The choices guiding my spontaneous movements, for instance typing a particular key ('r', 'a', 'd', 'o', etc.) while writing this paper, they originate from unobservable mental agency influencing the basic random dynamics of my brain.

The ``unobservable mental variable'' I introduce does not imply probabilities beyond the quantum ones, nor invalidates any law of Nature. The quantum philosophical view I propose overcomes the dualistic view of the soul ``as something divorced from the tangible grey matter''.

\subsubsection {``Why should an autonomous and self-conscious mind, directly created by God, need a brain to live and act in the material world?'' \cite{gr99}}

This objection overlooks an essential characteristic of human nature. Besides the impossibility of signaling faster than light (``no signaling''), there is another fundamental limitation defining the human condition: the impossibility of being in a \emph{permanent conscious state}. The human mind cannot be permanently aware of his own existence, the human will cannot act on purpose all the time. A human person is not a pure spiritual intellect (an angel), but a neuronal one, i.e. a mind who cannot be permanently self-conscious. A human brain is nothing other than the ensemble of conditions which make it possible that such a mind exists, basically through a sleep-wake cycle.

A brain is a device combining meaningless spontaneous outcomes and purposefully ordered ones, and a possibility to achieving this could come from the following \emph{rule}: At any moment a large series of brain outcomes must obey a statistical distribution deriving from the values of physiological parameters at this moment. Suppose that by producing wilful and meaningful pieces of information (short series of bits), the statistical distribution of the brain outcomes differs to some extent from the expected one. According to the \emph{rule} such a statistical deviation needs to be balanced by other periods (sleep, uncontrolled movements) restoring the large scale distribution.

I call this possibility the \emph{Quantum Homeostasis Hypothesis}. In particular, REM (Rapid Eye Movements) sleep may be important not only to the developing of the infants brain, but also to the homeostatic regulation of quantum distributions in each phase of life. REM denotes paradoxical stages of sleep, in which the brain is highly active \cite{ah}. REM sleep, and especially ``phasic REM'' \cite{rw}, may be part of the price we have to pay for the periods of intentional behavior and alertness during normal waking.

The \emph{Quantum Homeostasis Hypothesis} for the functionality of REM sleep is somewhat complementary to the \emph{Synaptic Homeostasis Hypothesis} for non-REM stages, which states that during sleep, the synapses weaken. This weakening restores the brain for the next period of learning: ``Sleep is the price we have to pay for plasticity, and its goal is the homeostatic regulation of the total synaptic weight impinging on neurons'' \cite{tc}.

In this perspective, \emph{no permanent conscious state}, similar to \emph{no signaling faster than light}, may provide a primitive principle for explaining why the laws of Nature are quantum.

My capacity for alertness is limited: I cannot keep driving a car indefinitely without sleeping; after a time I will begin to have random neuron firings, eventually hallucinate, and finally fall asleep at the wheel. Sleep may indicate when a series of signals from the brain-stem is ``large enough'' and must inescapably obey a statistical law imposed by physiological parameters. If so, this would be important information to complete quantum physics.

In any case, the \emph{Quantum Homeostasis Hypothesis} accounts well for: (a) the relevant fact that we experience ourselves as mental free beings capable of producing purposefully brain outcomes as speaking and acting, and (b) the available observations on REM sleep \cite{ao}. An important open question is why sleep proceeds in several cycles of alternating non-REM and REM stages, and why REM sleep itself exhibits cycles of tonic and phasic REM \cite{rw}.

\subsubsection {Consciousness, thinking, deciding are brain processes involving billions of neurons, i.e., macroscopic physical states, and therefore far away from quantum states \cite{gr99}.}

In the same line of thinking one states: ``Molecular machines, such as the light-amplifying components of photoreceptors, pre- and post-synaptic receptors and the voltage- and ligand-gated channel proteins that span cellular membranes and underpin neuronal excitability, are so large that they can be treated as classical objects. Although brains obey quantum mechanics, they do not seem to exploit any of its special features.''\cite{kh}

The objection originates from a widespread prejudice about the quantum. It overlooks that the decision about which detector clicks (in an interference experiment, Figure \ref{f1}) does not happen when ``one photon encounters a detector'' but only subsequently, after a \emph{virtual} cascade involving billions of electrons has been triggered. Only then an irreversible registration of a result happens and a human observer can become aware of it. In fact this means that the decision is not between ``one photon encountering D(+)'' and ``one photon encountering D($-$)'', but between ``a virtual assembly of electrons in D(+)'' and ``a virtual assembly of electrons in D($-$)''. The decision gives reality to one of these two virtual assemblies of electrons: A detection is an ``elementary act of creation'', in Anton Zeilinger's words.

The particular conditions defining \emph{when} precisely the decision happens are to date an unsolved (but solvable) problem (the measurement problem -see Objection 7 below). By contrast, as said above, the question of which detector clicks is a matter of an unobservable free decision, and as such cannot be answered. All this means that ``quantum effects'' (already at the level of simple interference experiments) consist in decisions about macroscopic outcomes occurring in visible classical objects (detectors).

Let us now consider a conscious decision about for instance moving the right or left hand. We know that this depends on the building of different ``transient neuronal assemblies'' \cite{sg}. The neuronal assemblies (like the counts in different detectors in quantum experiments) are measurable. But the choice between two rival neuronal assemblies, as the choice between two rival detectors, may very well originate from unobservable agency.

You can say that the difficulty in tackling this issue by experimental means is only due to the high inaccuracy of current measuring techniques: imaging techniques for instance are still too slow to capture the recruitment of ten million cells in less than a quarter of a second.

This was the kind of objection raised against indeterminism in the beginnings of quantum mechanics. The techniques for studying the neuronal activity will certainly improve, but not to the extreme of overcoming any indeterminacy. Just like information technology will not improve to the extreme of communicating faster than light.

If you assume that basically ``brains obey quantum mechanics'', and you are for freedom, then the reasonable attitude is to conclude that the realization of one specific neuronal assembly among several possible ones cannot be explained \emph{exclusively} through deterministic causality. As in the case of the detectors in the interference experiments, there is nothing in the observable universe, no story in the spacetime, capable of explain why this neuronal assembly happens and not another.

\subsubsection {There is no need for a metaphysical principle beyond the neurophysiological activity: Mind and consciousness can be considered physical states. \cite{gr99}}

In the same direction Wolf Singer compares the brain to ``an orchestra without conductor'': The brain is a highly complex system with nonlinear dynamics, and therefore capable of self-organization; the concept of a ``Self'' sharing immaterial operations is an illusion \cite{ws05}.

Singer's argument is rooted in a fallacy that comes from the intuitive way of thinking about causality. Intuition leads us to conceive time as the basis of causality and order. If we observe order and synchronization and cannot explain them by measurable causes in spacetime, we may intuitively be tempted to conclude that there is no cause at all, that ``No One'' acts through the brain. The argument has the merit of showing again that determinism leads to the loss of the personal identity.

If one assumes the freedom of the scientist as an axiom, then ``self-organization'' cannot be considered deterministic. Non-deterministic self-organization in a highly complex system like the brain can very well result from quantum effects amplified through nonlinear chaotic dynamics.

It is well known that quantum physics supports ``experimental metaphysics'': Nothing speaks against considering mind and consciousness \emph{quantum}-mechanical states of the brain. Actually, ``self-organization'' is another way of saying that random neural dynamics is controlled from outside spacetime by unobservable principles like free will and consciousness: ``Self-organization'' of the brain is synonymous to ``organization by the Self''.

\subsubsection {Specific molecular machines and proteins
have been proposed to implement quantum computations. However quantum computations are difficult to implement, and the brain does not seem to offer the conditions required to avoid decoherence \cite{kh}.}

Quantum computations are difficult to implement with a machine susceptible to utilization by any external technician. This requires very strong constraints to avoid decoherence. The brain is not such a machine: I can influence the randomness in my brain, but I cannot influence the randomness in your brain.

The efficacy of free will originates from influencing the order of the bits in a string of brain outcomes from outside spacetime, without violating the statistical distribution for a large series. This cannot be done by changing material variables, the same as in the experiment sketched in Figure \ref{f3} the order of the outcomes cannot be controlled by changing path-lengths.

\subsubsection {There is no evidence that consciousness is strictly necessary to the collapse of the wave function \cite{kh}.}

In fact, for measurement to happen it is not necessary at all that a human observer (conscious or not) is watching the apparatuses. However the very definition of measurement makes relation to human consciousness: An event is ``measured'', i.e. \emph{irreversibly} registered, only if it is possible for a human observer to become aware of it.

The concept of \emph{irreversibility} appears explicitly in the clinical definition of death, which basically states that death occurs when the neural functions responsible for certain spontaneous movements \emph{irreversibly} break down. In establishing death this way, we are assuming as obvious that our capacity of restoring neuronal dynamics (our repairing capability) is limited in principle, even if we don't yet know where this limitation comes from.

Similarly, one could assume that amplification in a photomultiplier becomes irreversible in principle at a certain level, if beyond this level an operation exceeding the human capabilities would be required to restore the photon's quantum state. When such a level is reached the detector clicks. Such a view combines the subjective and the objective interpretation of measurement: on the one hand no human observer has to be actually present in order that a registration takes place, just the same as in the GRW ``spontaneous collapse'' \cite{bg,rt}) or Penrose's ``objective reduction'' (OR) \cite{rp}; on the other hand one defines the 'collapse' or 'reduction' with relation to the capabilities of the human observer \cite{bd}.

Measurement is basic to quantum mechanics. But the theory does not define at all which conditions determine when measurement happens (measurement problem). This state of affairs clearly shows another point where the theory can and must be completed.

\section {Conclusion}

The preceding discussion has highlighted that free will is an axiom one can accept, or reject. The way scientists themselves behave overwhelmingly confirms the great acceptance of the axiom of free will: Any scientist will claim to be the conscious and free author of the work he publishes, and not a zombie repeating things already pre-determined in the Big-Bang. The scientist is part of the world. If he chooses to be free, then there is free will in the world.

\emph{Free will, though not permanently conscious,} is a primitive principle that implies a world sharing quantum features:

If you choose freedom, then you must coherently reject any \emph{pure} deterministic explanation of the world and in particular of the human brain and body.

The before-before experiment demonstrates that immaterial influences can control randomness to produce nonlocal patterns. Thus, immaterial free will can control the random dynamics of the brain to produce meaningful communication and behavior, but has to pay for it with uncontrolled dynamics during sleep (\emph{no permanent conscious state}). This constraint can also be formulated by stating that brain outcomes have to fulfil distributions imposed by neurophysiological parameters: statistical deviations after a certain amount of control (while waking) are corrected by further uncontrolled outcomes (while sleeping). Free will fits well with statistical laws of Nature like the quantum mechanical ones.

Conversely, if you choose a completely deterministic description of the brain, then you should coherently give up freedom and responsibility. You cannot even claim to be 'Someone' because determination by the past implies the loss of personal identity.

If you (like Wolf Singer) have a problem with these consequences, this may mean that you are not satisfied with your deterministic option, and you are actually wondering why a deterministic shaped universe produces guys like you and me who wish to be free.

\noindent\emph{Acknowledgments}: I am grateful to Roger Colbeck, Bernard d'Espagnat, Nicolas Gisin, Nick Herbert, and Valerio Scarani for comments.

\end{document}